\documentclass[12pt,]{article}
\usepackage{authblk}
\usepackage{fullpage}
\usepackage{amssymb,amsmath}
\usepackage[utf8x]{inputenc}
\usepackage[T1]{fontenc}
\usepackage{siunitx}
\usepackage[version=3]{mhchem}
\usepackage[ruled,vlined]{algorithm2e}
\usepackage{subcaption}
\usepackage{natbib}
\usepackage[left]{lineno}

\usepackage{setspace}
\usepackage[unicode=true]{hyperref}
\hypersetup{breaklinks=true,
            bookmarks=true,
            colorlinks=false,
            pdfborder={0 0 0}}
\usepackage{graphicx}
\graphicspath{{figures/}}
\makeatletter
\def\ScaleWidthIfNeeded{%
 \ifdim\Gin@nat@width>\linewidth
    \linewidth
  \else
    \Gin@nat@width
  \fi
}
\def\ScaleHeightIfNeeded{%
  \ifdim\Gin@nat@height>0.9\textheight
    0.9\textheight
  \else
    \Gin@nat@width
  \fi
}
\makeatother
\setkeys{Gin}{width=\ScaleWidthIfNeeded,height=\ScaleHeightIfNeeded,keepaspectratio}%

\title{COVID-19 Vaccine Hesitancy and Information Diffusion: An Agent-based Modeling Approach}\date{}

\author[1]{Pooria Taghizadeh Naderi}
\affil[1]{University of Tehran}

\author[2]{Ali Asgary}
\affil[2]{York University}

\author[2]{Jude Kong}

\author[2]{Jianhong Wu}

\author[1]{Fattaneh Taghiyareh}

\begin{document}

\maketitle

\newpage
\section*{Abstract}
Despite the unprecedented success in rapid development of several effective vaccines against the Cov-SARS-2, global vaccination rollout efforts suffer from vaccine distribution inequality and vaccine acceptance, leading to insufficient public immunity provided by the vaccine products. While a major current focus in vaccine acceptance research is how to model and inform vaccine acceptance based on social-demographic parameters, characteristics of vaccine acceptance are not well understood and in particular it is not known whether and how information diffusion influences vaccine acceptance. This study examines how information diffusion can change vaccine acceptance by developing a comprehensive computational model with agent-based simulation technique to overcome the modeling and quantification complexity associated with socio-demographics, vaccine types, population statistics, and information diffusion. Our analyses, calibrated by the vaccine acceptance survey data from the provinces and territories of Canada, provide clear evidence that propagation of information can greatly influence vaccine acceptance rates. The results illustrate that spread of negative messages about the COVID-19 vaccines can cause significant vaccine hesitancy that challenges the goal of a high public immunity provided by the vaccines. Our findings might help solve the vaccine hesitancy problem by focusing more on individuals' opinions and behavior.
\newpage
\section{Introduction}

Since the beginning of the COVID-19 pandemic, the vaccine has been considered a key solution to end the crisis. While many vaccine developers have been working towards developing effective vaccines, vaccine development and production alone are not enough to provide public immunity. Adequate number of people must be fully vaccinated to create immunity in the population. Vaccine hesitancy poses a challenge \cite{malik2020determinants,higgins2021looking,loomba2021measuring,bloom2020will,salali2020covid} and is increasingly becoming a vital factor in the vaccination process. There are intensive discussions about vaccine acceptance among policymakers, academic researchers, and the general public. Several studies, such as \cite{kwok2021influenza,gagneux2021intention,deroo2020planning,detoc2020intention}, have assessed vaccine hesitancy in nations, and it is apparent that there is a considerable unwillingness in different countries that may negatively impact the vaccination rate for the Cov-SARS-2. Unlike the previous pandemics and diseases, information and misinformation about vaccines are spread rapidly among the public through mass media and social media during the COVID-19 pandemic. Therefore, understanding the flow of information within  social media platforms can help predicting how a product is accepted or refused in the society and what elements can promote its acceptance or wider use \cite{9443155}. 

An individual's decision to be vaccinated or not is influenced by personal, social, and political factors, including information received from family members, friends, peers and users of online social networks. In order to develop effective vaccination strategies and interventions, it is critical to learn how social influences impact vaccine acceptance. Several factors such as  vaccine safety, disease severity, and the vaccine price play a key role in an individual's vaccination decision. However, few studies have been conducted to assess how individuals' beliefs and social ties influence vaccine acceptance. In particular, individuals' beliefs may change after exposure to feedback and comments from those family members, friends, coworkers and other users sharing the social media platform who have received the vaccine.

The main goal of this study is to examine how different factors, particularly social influences, impact individuals' decisions to take a vaccine and vaccine propagation in a community using agent-based modeling and analysis. We believe that we have designed an innovative simulation toolkit to study vaccine acceptance that can be used in different societies. In our platform, we have incorporated mechanisms to quantify the importance of information diffusion. Our research and simulation tool could be a useful aid for decision-makers because they give them a reliable platform to examine and test their decisions. using the simulation tool, it is possible to measure how promoting positive messages through social media, advertisements, and national media campaigns might increase vaccine acceptance in the region or the country.

\section{Background and literature review}

In response to the COVID-19 pandemic, many studies have been conducted to understand vaccine acceptance, vaccine hesitancy, and vaccine social influence. Based on our literature review, most of these studies focus on COVID-19 vaccine acceptance in general (28 articles), followed by much empirical research on COVID-19 vaccine acceptance in a specific country (16 articles). Very few studies have used modeling and simulation approaches for examining vaccine acceptance and propagation (3 articles).
\subsection{Vaccine acceptance in general}
Empirical research results show that despite vaccine availability, vaccination hesitancy has typically been observed uncontrolled and varies widely among different countries \cite{sallam2021covid}. UK residents, for instance, displayed the lowest hesitancy level of 10.1\%, while Kuwaitis exhibited a very high hesitancy level of 76.4\% \cite{sallam2021covid,lazarus2021global}. These studies revealed that individuals' characteristics, including age, gender, income, and education contribute to vaccine hesitancy \cite{murphy2021psychological,daly2021willingness}. Other studies suggested that in the early period of vaccine availability in the market, its hesitancy was much higher \cite{fadda2020covid}. The study \cite{latkin2021trust} also showed that a person's trust in the COVID-19 vaccine and factors contributing to mistrust may play a role in vaccine acceptance in different societies. According to Latkin et al. \cite{latkin2021trust}, trust is low when people believe or perceive that vaccines were being distributed too soon without adequate testing, data, or proof of success. Mistrust is also linked with uncertainty and lack of knowledge about the long-term side effects, and people who adhere to natural immunity distrust vaccines as a whole or do not like them. According to \cite{latkin2021trust}, there are also concerns about the efficacy of the COVID-19 vaccines. The study  \cite{callaghan2021correlates} concluded that individuals who do not plan to get vaccinated against COVID-19 do not believe that the vaccine will be safe or effective. Other less significant reasons for vaccine refusal include lack of insurance or financial resources. Several groups believe that they have already contracted COVID-19 and therefore refuse to take vaccine. This study \cite{callaghan2021correlates}  also confirmed that women and African Americans have greater hesitancy regarding COVID-19 vaccination, and they explore the reasons why women and African Americans are less apt to refuse vaccination. It has been reported that women did not pursue vaccination as often as men because they did not consider the vaccine safe or believed that it would not be effective. Moreover, based upon the statistical analysis of differences among races, blacks are more likely than whites to decline vaccination for reasons included in their survey, with statistical signals for safety concerns, effectiveness, inadequate insurance, and financial resources. Another aspect of vaccine acceptance that has not been studied in the past is the average vaccination time. In a survey \cite{muqattash2020survey}, participants were asked about vaccination  time and its effect on vaccination decisions, and the results indicated that average vaccination times affect vaccine acceptance rates. Studies showed that vaccine prices play a significant role in vaccine hesitancy, and in some countries, free vaccination is the result \cite{garcia2020contingent,yin2021unfolding,muqattash2020survey}. \cite{garcia2020contingent} aimed to estimate the individual's willingness to pay (WTP) in response to a hypothetical COVID-19 vaccine. This study determined that the vaccine's price could affect vaccine acceptance in the society. Their results concur with prior research, which found that the average family income could affect vaccine resistance. Many people believe that the government should finance the vaccine.

\subsection{Misinformation}
Most studies have focused on individual characteristics like gender, race, education, job, and vaccine attributes like price and expiry dates. However, another crucial aspect that has received less attention is the misinformation that participants receive through social media or their family members and friends. A research by García et al. \cite{garcia2020contingent} involved a randomized controlled trial in the UK and the USA to determine how online misinformation on COVID-19 vaccines influences individuals' intent to protect themselves and others. They asked participants to fill a survey and answer questions related to their intention before being exposed to misinformation. Individuals intent to get vaccinated is influenced by misinformation based on different socioeconomic factors. For example, the UK results showed that unemployed persons are more susceptible to misinformation than employed individuals. Further, whites and lower-income groups in the USA are more vulnerable to misinformation than other racial and ethnic groups. Surprisingly, trusting celebrities is associated with greater resilience to misinformation in the UK than controls. Conversely, trusting family members and friends in the USA is associated with lower resistance to misinformation. The study\cite{chou2020considering} discussed several methods of leveraging emotion in vaccine communication for the COVID-19 vaccines, both in the short term to address vaccine hesitancy and in the long term to promote vaccine confidence. They discovered that misinformation about vaccines could be eliminated by making society aware of how misinformation campaigns use negative emotions and beliefs.

The effects of misinformation campaigns on vaccination intentions have been documented in \cite{loomba2021measuring}. However, they have not seemed proportional to how information has reached a person in social media platforms as a complex mix of what media has shown and what others have shared. In the current environment of online social network structures, which result in homophilic social interaction, selective exposure and creation of echo chambers or polarization of opinions can amplify or dampen the spread of misinformation. Current studies introduce many different aspects of vaccine hesitancy which we summarize them in Table \ref{table: 1}. However, they were limited by the fact that they do not create a detailed model that would demonstrate similar behaviors to what would occur in the real world in the case of the COVID-19 pandemic.

\begin{table*}[t]
  \centering
  \caption{A snapshot of the main influences on vaccination decision-making among individuals}
  \label{table: 1}
  \begin{tabular}{p{0.6\linewidth} | p{0.3\linewidth}}

  \hline
 Hesitancy reasons & Reference \\ [0.5ex] 
 \hline\hline
 Too new, worried about unforeseen effects, Vaccine fast-tracking & \cite{latkin2021trust,paul2021attitudes} \\
 \hline
 Compromised immune systems, preference for natural immunity & \cite{latkin2021trust,paul2021attitudes} \\
 \hline
 Profit distrust, concerns about commercial profiteering & \cite{latkin2021trust,paul2021attitudes} \\
 \hline
 Current government distrust & \cite{latkin2021trust} \\
 \hline
 Vaccine contents, Vaccine skepticism, Vaccine refusal in general, General mistrust of vaccine benefits, General skepticism, Vaccine being unnecessary, trust in research and in vaccines & \cite{latkin2021trust,ni2021towards,goldman2020caregiver,paul2021attitudes,palamenghi2020mistrust} \\
 \hline
 Virus strain/mutation & \cite{latkin2021trust,loomba2021measuring} \\
 \hline
 Doubt in efficacy, Doubt in efficacy, Vaccine Won’t be Effective & \cite{latkin2021trust,goldman2020caregiver,callaghan2021correlates} \\
 \hline
 Side effects, Side effects/safety concerns, Vaccine Won’t be Safe, Perceived Contraindication & \cite{ni2021towards,goldman2020caregiver,callaghan2021correlates} \\
 \hline
 Perceived low risk of disease, Perceived child not at risk to contract COVID19, Already had COVID-19 & \cite{ni2021towards,goldman2020caregiver,callaghan2021correlates} \\
 \hline
 Difficulty or inconvenience in accessing vaccination services & \cite{ni2021towards} \\
 \hline
 Vaccinate if more information available & \cite{goldman2020caregiver} \\
 \hline
 Lack of Insurance, Lack of Financial Resources & \cite{callaghan2021correlates} \\
    \hline
  \end{tabular}
\end{table*}

\section{Materials and Method}

We develop an agent-based model to describe the vaccine acceptance and the influence of communication, vaccine characteristics, and socio-demographic attributes of individuals. We use AnyLogic Software version 8.7.3. We simulate if and when people accept vaccines and the model assesses the impacts of different communication channels on people’s vaccine acceptance behavior. Simulation can run for any period to capture the dynamics of vaccine acceptance over time and demonstrate the latest information about vaccine and disease spread. The simulation has been developed to be run for any country given the initial parameter values is available at country and regional levels within a country. In this paper we apply the model to Canada and its 13 provinces and territories to demonstrate the simulation results. Our agent based model consists of four agent types: Person, Country, Vaccine and Communication.

\subsection{Person Agent}

\begin{figure}[t]
    \noindent
    \includegraphics[width=\textwidth]{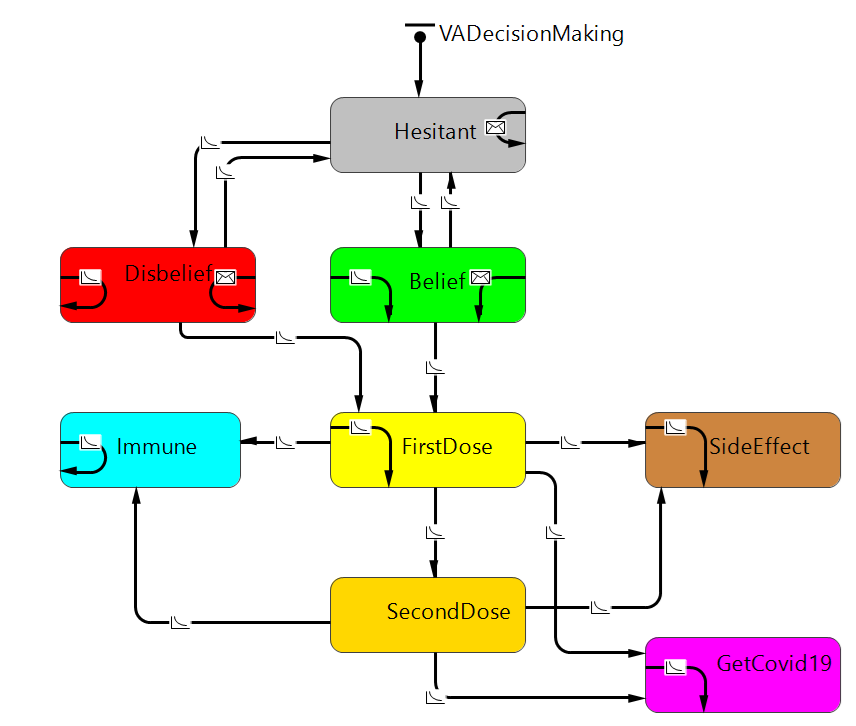}
    \caption{Person agent state chart}
    \label{figure: 1}
\end{figure}

The person agent is the major agent in this ABM. Figure \ref{figure: 1} shows the main statechart for the Person agent. Once a person agent is created a logistic regression function runs that indicates the initial vaccine acceptance of the person based on the agent’s attributes such as age, gender, education, income, and race (Table \ref{table: 2}). The logistic regression (Eq.\ref{equation: 1}) used in this study is based on the previous studies on vaccine hesitancy in different countries and regions \cite{loomba2021measuring,lazarus2021global,borriello2021preferences,alqudeimat2021acceptance}. For each demographic feature, $OR_i$ indicates the odds ratio, while CO represents the coefficient value that represents the sum of all $OR_i$, and P stands for vaccine acceptance probability.

\begin{equation}\label{equation: 1}
CO = \sum \log (OR_i)\\
P = \frac{e^{CO}}{{e^{CO}} + 1}
\end{equation}

\begin{algorithm}
\DontPrintSemicolon
\SetAlgoLined
\KwResult{Next state }

$P,N,Opinion = 0$\;
 \While{ $state \in \{Belief,Disbelief\}$ }{
  m = new message\;
  \eIf{m > 0}{
   P = P + 1\;
   }{
   N = N + 1\;
  }
  $Opinion = \frac{P}{P+N}$\;
  \eIf{$state=Belief$ and $Opinion < \theta$}
  {\Return Hesitant}{
  \eIf{$state=Disbelief$ and $Opinion > \theta$}
  {\Return Hesitant}{
  Continue
  }
  }
 }
 \Return Belief
 
 \caption{Information diffusion algorithm for disbelief state}\label{alg: 1}
\end{algorithm}

We assume that all persons are in a hesitant state with the vaccine at the beginning. Their status changes to either Disbelief or Belief states based on the probability we calculated in Eq.\ref{equation: 1}. A person’s Hesitant, Disbelief and Belief states are influenced by the information received from different sources including friends, media and government. In order to measure positive and negative information diffusion, we used a threshold-based algorithm \cite{singh2018survey}, which counts the number of positive and negative information users may receive in each step. Based on the algorithm \ref{alg: 1}, we decide to change the user state from the Belief or the Disbelief to Hesitant at an appropriate rate. This algorithm requires two input parameters, Teta and InfoEffect. Teta is used to control when a person changes his/her state, and InfoEffect measures how quickly an agent moves from the Belief state or the Disbelief state to the Hesitant state. For example, in the case where Teta = 0.3, you may see that the person changes his/her state from the Disbelief to the Hesitant if the number of positive messages exceeds 30 percent of the total messages. A person that is in the Belief state is going to take the vaccine (first dose in cases of two doses vaccine) when it becomes available to her or him. Also, with the addition of the Vaccine Passport (or certificate) parameter to the Person agent, the vaccine passport can now be considered mandatory. Regardless of being in Disbelief, in this case, if the vaccine passport parameter is set to 1, person agents would consider getting the vaccine. If the person experiences side effects, he/she may send negative information to others that may influence their vaccine acceptance level. If the person becomes infected after the first or second dose, he or she may send negative information to others. On the other hand, if this person becomes immune, he or she will send a positive message to others. We assume that the information diffusion between individuals that are connected can change vaccine acceptance in the society. In the simulation tool, users can set the parameters according to their study area information and then investigate different scenarios. 

\begin{table*}[t]
  \centering
  \caption{Summary of Agent person parameters}
  \label{table: 2}
  \begin{tabular}{|p{0.27\linewidth} | p{0.36\linewidth}| p{0.27\linewidth}|}
  \hline
  Person agent attributes & Description	& Attribute values/range\\
  \hline\hline
  Age & Age of the person & 18-29; 30-39; 40-49;50+\\
  \hline
  Sex & Sex of the person & Male, Female\\
  \hline
  Education & Education level of the person & High school and less; University/college degree\\
  \hline 
  Income & Income of the person & Low; Medium; High\\
  \hline
  Race & Racial background of the person & White, black, Asian, other\\
  \hline
  Contact rate & Average of daily contacts between the person and others & 1\\
  \hline 
  Vaccine passport  & Whether having vaccine passport is used in the country or not & Binary, 0, 1\\
  \hline
  Vacacine passport effect rate & How much effective is vaccine passport rule & Vaccination speed in the country /10\\
  \hline
  DedicatedVaccine Type & Type of vaccine provided to the person & E.g. Pfizer, Moderna\\
  \hline
  Acceptance & Vaccine acceptance level by the person & 0-1\\
 \hline
  \end{tabular}

\end{table*}

% \begin{figure}
%     \includegraphics{fig1.png}
%     \caption{Person agent state chart}
%     \label{figure: 1}
% \end{figure}

\subsection{Province agent}

The country/Province agent is linked with the person agent and holds the demographic, vaccination, COVID-19, and vaccine hesitancy information. Custom distributions are used for age, education, gender, race, income, and vaccine distribution. All parameter values are extracted from a database that can be updated as new information becomes available. The agent province parameters consist of ten items. The following operators made up Equation \ref{equation: 1}: Sex OR, Education OR, Income OR, and Race OR. Furthermore, we created a random network between agents using Link Per Agent and Population Total to simulate real-life connections between agents. Additionally, we explored information diffusion using Teta, the InfoEffect, and Media Effects. A GIS map is used to demonstrate the vaccine %acceptance in each province visually (Figure.\ref{figure: 2}). 
Person agents are living in the province agent and thus receive relevant attributes from their host province.

% \begin{figure}[t]
%     \noindent
%     \includegraphics[width=\textwidth]{fig2.png}
% 	\caption{This is an example of a province dashboard with a GIS map for Alberta in Canada.}
% 	\label{figure: 2}
% \end{figure}

\subsection{Vaccine agent}
Vaccine agent represents vaccine and its attributes and are linked with the Person and Country agents. Vaccine parameters are also stored in the database and are updated as new information about vaccines become available. Vaccine parameters include name, manufacturer, efficacy, origin, number of doses required, development approval, type, and side effects. In each case, different types of vaccines are distributed among individuals with a different percentage. In Canada, Pfizer, Moderna, and Oxford/AstraZeneca vaccines are mostly used. Nevertheless, the data on all vaccines are available in the model database and \ref{covidTracker} if they need to be considered. 
% \begin{table*}[t]
%   \centering
%   \caption{Summary of Agent vaccine parameters and type}
%   \label{table: 3}
%   \begin{tabular}{|c||c|c|c|c|c|c|}
%   \hline
%   Name & Company & Efficacy & Company/Origin & Doses	Development & Approval & Type\\
%   \hline\hline
%   Comirnaty	& Pfizer & 0.913 & USA & 2 & Many Approved & mRNA\\
%   \hline
%   mRNA-1273 & Moderna & 0.9 & USA & 2 & One & mRNA\\
%   \hline
%   Ad26.COV2.S & JohnsonAndJohnson & 0.68 & USA & 1 & Paused & Ad26\\
%   \hline
%   Sputnik V & Gamaleya & 0.916 & Russia & 2 & Early & Ad26\\
%   \hline
%   Vaxzevria & OxfordAstraZeneca & 0.76 & UK & 2 & Stopped & ChAdOx1\\
%   \hline
%   Convidecia & CanSino & 0.6528 & China & 1 & One & Ad5\\
%   \hline
%   EpiVacCorona & VectorInstitute & 0.5 & Russia & 2 & Early & Protein\\
%   \hline
%   NVX-CoV2373 & Novavax & 0.96 & USA & 2 & Not Approved & Protein\\
%   \hline
%   BBIBP-CorV & Sinopharm & 0.7934 & China & 2 & One & Inactivated\\
%   \hline
%   CoronaVac & Sinovac & 0.506 & China & 2 & One & Inactivated\\
%   \hline
%   Covaxin & BharatBiotech & 0.78 & India & 2 & Not Approved & Inactivated\\ [0.5ex]
%  \hline
%   \end{tabular}
% \end{table*}

\subsection{Communication agent}

The communication agent is essential in this model. 
%Figure \ref{figure: 3} presents the state chart for the vaccine Communication agent. 
We assume that the Person agent's vaccine acceptance is influenced by different communication agents, including Radio, Newspapers, Television, Social media, Government Agencies, and Health Professionals. Based on the simulation, a communication agent can either be active or inactive. Additionally, an agent can send positive or negative messages, affecting individual users based on the Effect Rate parameter, and become dormant for some time. Various communication agents have different effect rates, and we considered this in our Communication agent. For instance, social media can have a more significant influence than radio.  

% \begin{figure}[t]
%     \noindent
%     \includegraphics[width=\textwidth]{fig3.png}
% 	\caption{Communication agent state chart}
% 	\label{figure: 3}
% \end{figure}

\subsection{Experiment setup}

As our model can be applied to any region with a population, we selected Canadian provinces and territories to validate it. We need to provide data such as vaccination type distributions and population distributions in order to run the simulation. As a primary distribution for Canada, we used \cite{covidTracker} and \cite{statcan}. In order to calculate individuals' acceptance of the vaccine, we needed an odd ratio of vaccine acceptance or hesitancy, so we used \cite{lunsky2021beliefs} as the input for the logistic function.
%(Table \ref{table: 6}). 
Based on reported vaccination data in Canada, we calibrated Dose 1 speed, Dose 2 speed, and Information Effect. Next, in a Monte Carlo simulation, we determined how the stochastic component of the model can impact the outcome. Based on reported vaccination data in Canada, we calibrated Dose 1 speed, Dose 2 speed, and Information Effect. This data consists of 144 rows and Date, Total Vaccination, and At least One Dose column. Each row stands for one day; total vaccination presents the cumulative number of fully vaccinated individuals, and At least One Dose is the sum of the first dose and second dose until that day. Next, in a Monte Carlo simulation, we determined how the stochastic component of the model can impact the outcome.

\subsection{Calibration}
As previously mentioned, the model aims to simulate vaccine acceptance, so that the total vaccination percentage serves as a comparison parameter between the simulation and the actual data. We compared our total vaccination rate per day for each province and territory in Canada by calibrating first dose speed, second dose speed, and InfoEffectRate on VaccineTracker \cite{covidTracker} data. These features were calibrated using Anylogic, as shown in Table \ref{table: 7}. It appears that the second dose speed is considerably lower than the first, leading to a relatively insufficient level of vaccination in Canada.

\begin{table*}[t]
  \centering
  \caption{Model Calibration for Canada total vaccination}
  \label{table: 7}
  \begin{tabular}{|c|c|c|c|}
  \hline
  Province & Info Effect Rate & First Dose speed & Second Dose Speed\\
  \hline\hline
  British Columbia & 0.2 & 0.033 & 3.08E-4\\
  \hline
  Alberta & 0.2 & 0.066 & 3.606E-4\\
  \hline
  Saskatchewan & 0.2 & 0.066 & 3.606E-4\\
  \hline
  Manitoba & 0.2 & 0.066 & 3.606E-4\\
  \hline
  Ontario & 0.2 & 0.066 & 3.606E-4\\
  \hline
  Quebec & 0.4 & 0.014 & 1.776E-4\\
  \hline
  New Brunswick & 0.2 & 0.066 & 3.606E-4\\
  \hline
  Nova Scotia & 0.2 & 0.061 & 3.588E-4\\
  \hline
  Prince Edward Island & 0.2 & 0.072 & 3.66E-4\\
  \hline
  Newfoundland and Labrador & 0.2 & 0.028 & 3.063E-4\\
  \hline
  Yukon & 0.2 & 0.018 & 0.004\\
  \hline
  Northwest Territories & 0.2 & 0.06 & 0.005\\
  \hline
  Nunavut & 0.2 & 0.046 & 0.004\\ [0.5ex]
 \hline
  \end{tabular}

\end{table*}

\section{Results}

\begin{figure}[t]
    \noindent
    \includegraphics[width=\textwidth]{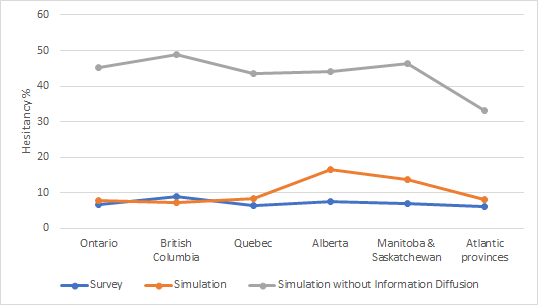}
	\caption{A comparison of hesitancy rate between the simulation and ground-truth data}
	\label{figure: 4}
\end{figure}

This paper examines how to model vaccine acceptance based on various factors, including socio-demographics, vaccine types, and information diffusion. To begin the simulation, we determined the vaccine acceptance based on an individual's socio$-$demographic characteristics. Next, we looked at how information propagation and media exposure affect vaccine acceptance. Therefore, our initialization method must be validated by comparing our results with similar surveys conducted in Canada. Generally, our preliminary results indicate that vaccine acceptance is similar to surveys and information diffusion parameters significantly affect vaccine acceptance.
As we enroll all users with a hesitation state, we can measure hesitancy for each province by counting agents in the Disbelief state (Figure.\ref{figure: 1}) after 90 days. Since March 1st is the first day in our dataset\cite{covidTracker} and the survey\cite{tang2021quantifying} was conducted between January and March 2021, we waited 90 days. We show that the vaccine hesitancy among agents is similar to the data collected in Canada, suggesting that the logistic function implemented works as expected. The simulation was also performed without the information diffusion effect to illustrate the differences in outcomes.
Figure.\ref{figure: 4} illustrates that the simulation results are comparable to the surveys with a RMSE of 0.802, and similar behavior was observed in all provinces and territories, with no sudden or unexpected changes. It can be observed from Figure.\ref{figure: 4} that there is a significant gap in Alberta between survey and simulation because we use 4,472 agents for the Alberta population, but their sample size is 301 for Alberta, which leads to these differences. Clearly, the information diffusion removal (gray line) increases the error and strongly confirms the effect of information diffusion on vaccine acceptance. Our comparison was limited to one similar research study about the hesitancy/acceptance in Canadian provinces \cite{tang2021quantifying}, therefore, it cannot be directly compared with other surveys.

\subsection{Vaccine hesitancy and socio$-$demographic}

\begin{figure}[t]
    \noindent
    \includegraphics[width=\textwidth]{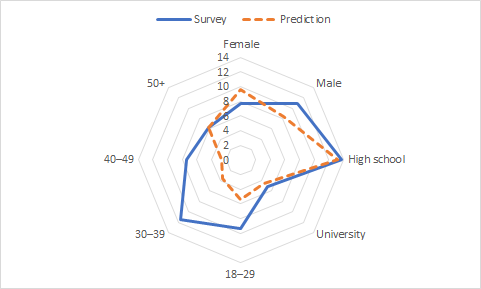}
	\caption{A comparison of hesitancy rate between the simulation and ground-truth data in different categories of socio-demographic variables}
	\label{figure: 5}
\end{figure}

In order to make sure our simulation does not have any specific bias, we decided to analyze the acceptance of vaccines in each category of provinces and territories based on the socio-demographic characteristics. We calculated hesitancy percentages for each group and compared them with data from the previous study\cite{tang2021quantifying}. Figure \ref{figure: 5} shows that the model estimation matches the reported data by \cite{tang2021quantifying}, and the RMSE for each demographic attribute is less than 5.06. In most cases, such as the University, High school, Female, and Male, simulation revealed a similar hesitancy percentage. Age categories' hesitancy percentage is slightly lower than the value we expected, and there is certainly room for improvement. It should be noted that we were unable to find any data regarding race and income in Canada. 

\begin{figure*}
     \centering
     \begin{subfigure}[b]{0.3\textwidth}
         \centering
         \includegraphics[width=\textwidth]{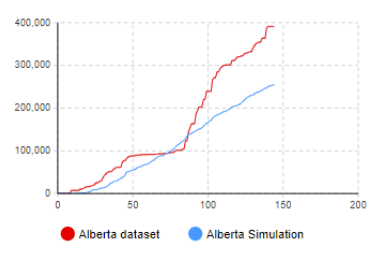}
         \caption{$y=x$}
         \label{fig:6-a}
     \end{subfigure}
     \hfill
     \begin{subfigure}[b]{0.3\textwidth}
         \centering
         \includegraphics[width=\textwidth]{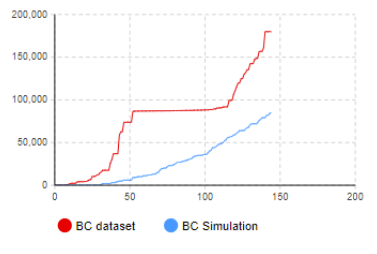}
         \caption{$y=3sinx$}
         \label{fig:6-b}
     \end{subfigure}
     \hfill
     \begin{subfigure}[b]{0.3\textwidth}
         \centering
         \includegraphics[width=\textwidth]{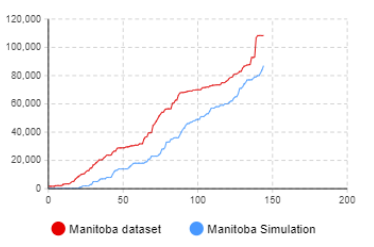}
         \caption{$y=5/x$}
         \label{fig:6-c}
     \end{subfigure}
     \begin{subfigure}[b]{0.3\textwidth}
         \centering
         \includegraphics[width=\textwidth]{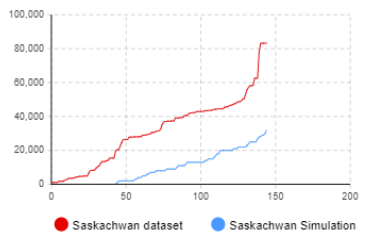}
         \caption{$y=5/x$}
         \label{fig:6-d}
     \end{subfigure}
     \begin{subfigure}[b]{0.3\textwidth}
         \centering
         \includegraphics[width=\textwidth]{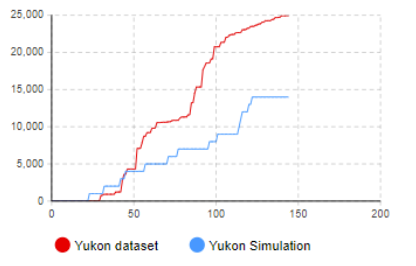}
         \caption{$y=5/x$}
         \label{fig:6-e}
     \end{subfigure}
     \begin{subfigure}[b]{0.3\textwidth}
         \centering
         \includegraphics[width=\textwidth]{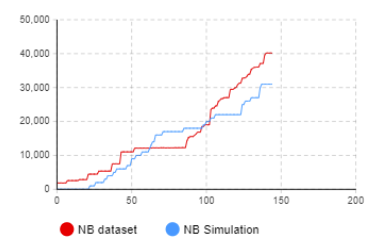}
         \caption{$y=5/x$}
         \label{fig:6-f}
     \end{subfigure}
     \begin{subfigure}[b]{0.3\textwidth}
         \centering
         \includegraphics[width=\textwidth]{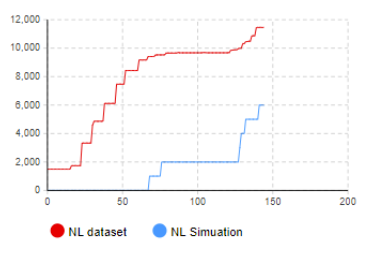}
         \caption{$y=5/x$}
         \label{fig:6-g}
     \end{subfigure}
     \begin{subfigure}[b]{0.3\textwidth}
         \centering
         \includegraphics[width=\textwidth]{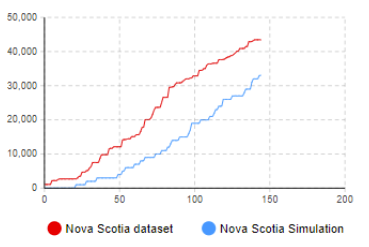}
         \caption{$y=5/x$}
         \label{fig:6-h}
     \end{subfigure}
     \begin{subfigure}[b]{0.3\textwidth}
         \centering
         \includegraphics[width=\textwidth]{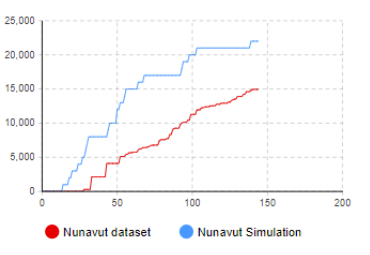}
         \caption{$y=5/x$}
         \label{fig:6-i}
     \end{subfigure}
     \begin{subfigure}[b]{0.3\textwidth}
         \centering
         \includegraphics[width=\textwidth]{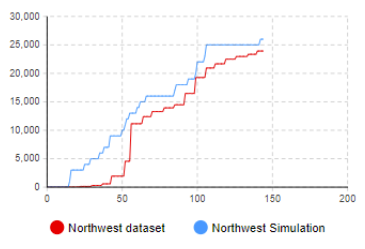}
         \caption{$y=5/x$}
         \label{fig:6-j}
     \end{subfigure}
     \begin{subfigure}[b]{0.3\textwidth}
         \centering
         \includegraphics[width=\textwidth]{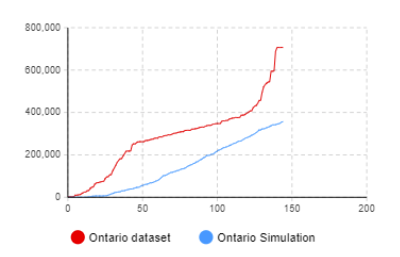}
         \caption{$y=5/x$}
         \label{fig:6-k}
     \end{subfigure}
     \begin{subfigure}[b]{0.3\textwidth}
         \centering
         \includegraphics[width=\textwidth]{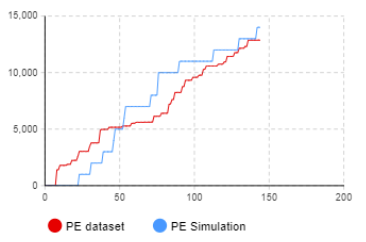}
         \caption{$y=5/x$}
         \label{fig:6-l}
     \end{subfigure}
     \begin{subfigure}[b]{0.3\textwidth}
         \centering
         \includegraphics[width=\textwidth]{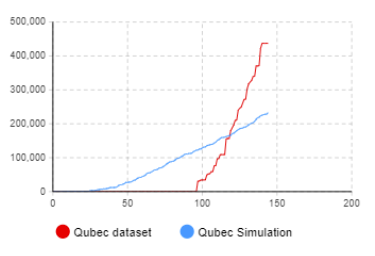}
         \caption{$y=5/x$}
         \label{fig:6-m}
     \end{subfigure}
        \caption{Predicting full vaccination in comparison with actual full vaccination in Canada's provinces}
        \label{figure: 6}
\end{figure*}

\subsection{Fully Vaccinated prediction (calibration)}

It is crucial to compare the speed of vaccination in Canada with our simulation results because this work aimed to test the effect of information diffusion on vaccination. A Canada vaccination report dataset \cite{covidTracker} consists of two columns: Total Vaccination and Fully Vaccinated. All of this data is manually compiled from credible sources in near real-time as new information becomes available. Despite 315 days of Canada vaccination data, we chose to run our model for 144 days since the last 144 days had a non-zero value for the fully vaccinated population. We used this dataset and the Anylogic calibration method to tune the Dose1Speed and the Dose2Speed parameters of our model. Figure \ref{figure: 6} indicates a good correlation between our model and vaccination rates across provinces. It can be observed from Figure \ref{figure: 6} that the patterns of simulation and real-world data are essentially the same in NB, PE, Northwest, and Manitoba. Similar behavior was observed in the majority of cases, with no sudden changes. However, the simulation underestimated the number of fully vaccinated individuals in NL and Quebec, but it follows the dataset pattern. The Quebec result(Figure \ref{fig:6-m}) is slightly lower than the dataset value, and we think the reason is that the Quebec dataset is not updated every day.

% It seems highly probable that the Quebec dataset is not updated every day, and they reported the aggregated data after 90 days. It appears that this problem caused our simulation did not follow the pattern of Quebec very well.

\begin{figure}
     \centering
     \begin{subfigure}[b]{0.45\textwidth}
         \centering
         \includegraphics[width=\textwidth]{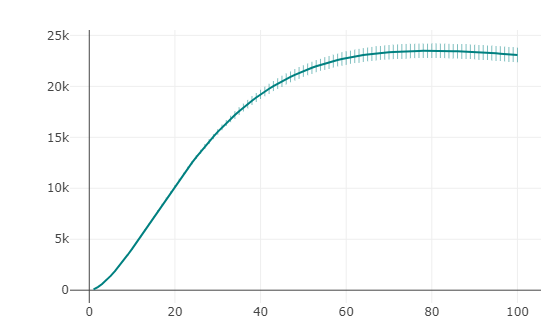}
         \caption{x=Days , y= Number of injections}
         \label{fig:7-a}
     \end{subfigure}
     \hfill
     \begin{subfigure}[b]{0.45\textwidth}
         \centering
         \includegraphics[width=\textwidth]{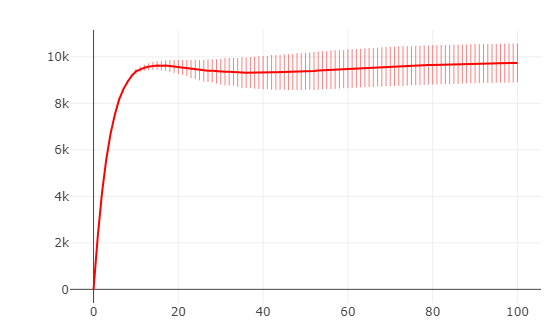}
         \caption{x=Days , y= Number of users in thedisbelief state}
         \label{fig:7-b}
     \end{subfigure}
     \hfill
        \caption{A robustness test of the simulation with monte-Carlo. \textbf{(a)} The number of agents in the Immune state on a scale of 1000. \textbf{(b)} The number of Agents in the Disbelief state on a scale of 1000}
        \label{figure: 7}
\end{figure}

\begin{figure}
     \centering
     \begin{subfigure}[b]{0.45\textwidth}
         \centering
         \includegraphics[width=\textwidth]{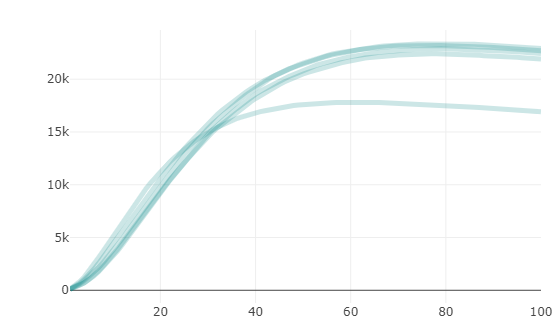}
         \caption{x=Days , y=Number of injections, Teta is varied}
         \label{figure: 8-a}
     \end{subfigure}
     \hfill
     \begin{subfigure}[b]{0.45\textwidth}
         \centering
         \includegraphics[width=\textwidth]{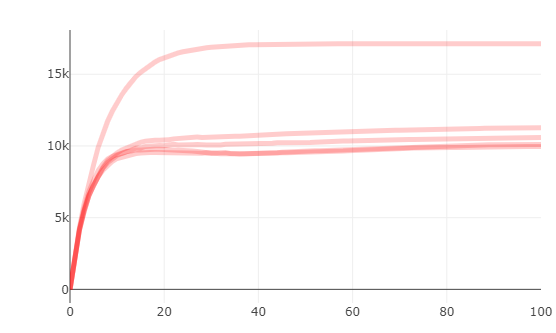}
         \caption{x=Days , y=Number of users in the disbelief state, Teta is varied}
         \label{figure: 8-b}
     \end{subfigure}
     \hfill
     \begin{subfigure}[b]{0.45\textwidth}
         \centering
         \includegraphics[width=\textwidth]{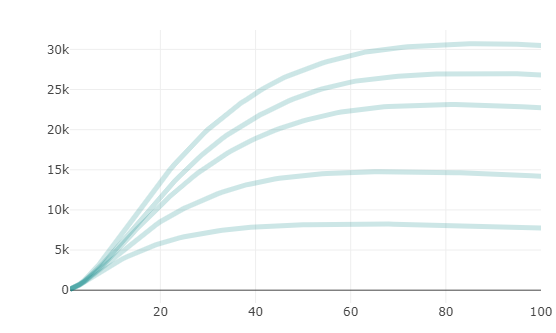}
         \caption{x=Days , y=Number of injections, TInfoEffect is varied}
         \label{figure: 8-c}
     \end{subfigure}
     \begin{subfigure}[b]{0.45\textwidth}
         \centering
         \includegraphics[width=\textwidth]{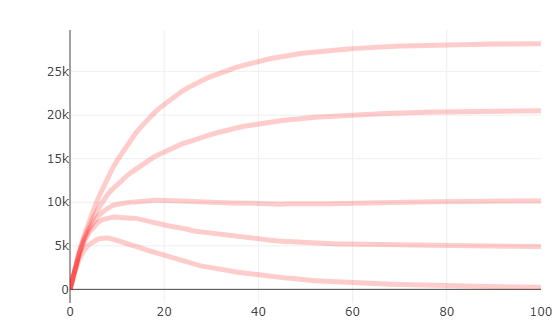}
         \caption{x=Days , y=Number of users in the disbelief state, InfoEffect is varied}
         \label{figure: 8-d}
     \end{subfigure}
        \caption{Teta \textbf{(a,b)} and InfoEffect \textbf{(c,d)} varied from zero to one to compare the immunity rate in Canada simulation. \textbf{(a,c)} The number of agents in the Immune state on a scale of 1000. \textbf{(b,d)} The number of Agents in the Disbelief state on a scale of 1000.}
        \label{figure: 8}
\end{figure}

\subsection{Monte Carlo - User states}
According to the Monte Carlo experiment, the results indicate that uncertainty did not significantly impact the model's behavior (see Figure \ref{figure: 7}.) For each iteration of the Monte-Carlo simulation, we counted the individuals in each state. Based on the results of the Monte-Carlo experiment, it is evident that agents' behavior is affected by information effects, Dose1Speed, and Dose2Speed parameters despite the stochastic nature of the simulation. The results of other simulations that are not part of the Monte-Carlo simulation are also trustworthy, and the results of the model run multiple times will say the same.

\subsection{Information Diffusion Effects}

As mentioned early, our primary goal in this paper is to  construct a computational platform and use it to simulate and analyze  the effect of information propagation on vaccine acceptance. Therefore, we compared the number of people who received at least one dose vaccine according to different scenarios by varying the  Teta and InfoEffect parameters. Generally speaking, both parameters resulted in a change in vaccination rates. First, we fixed all parameters except the InfoEffect, and then we increased its value from zero to one by 0.25 each step, and we repeated the same scenario for the Teta parameter. As listed in Figure\ref{figure: 8}, this parameter affects the vaccination process as expected. When the Info Effect is equal to zero, the number of vaccinated people is entirely different from the real-world data. The population that has received at least one dose for Canada is nearly 26 million, and our simulation returns the same number when the Info Effect is not zero. However, when we set the Info Effect equal to zero, we get nearly 15 million. Comparing Figure\ref{figure: 8-c} and Figure \ref{figure: 8-d} shows that the number of agents in Disbelief states is much higher when we ignore the Info Effect parameter. This result proves that the information propagation could change people's minds and encourage them to get vaccinated. We observe from Figure\ref{figure: 8-a},\ref{figure: 8-b} that the Teta parameter also affects the vaccination rate similar to the Info Effect parameters. This result confirms previous findings and shows that the Teta value could considerably change the outcomes. The number of vaccinated agents increased to nearly 30 million after setting Teta to 0 and relaxing our model. Conversely, the vaccinated agents are close to 5 million when the Teta is 1.

\section{Discussion and Conclusion}

As outlined in the introduction, previous research simulated the vaccination focused on parameters such as socio-demographics or vaccine types, but they ignored the effect of information diffusion. Through agent-based modeling, this study examined how various factors, such as social influences, affect vaccination decisions. We defined Person, Province, Communication and Vaccine agent to model the vaccination rate. This model can be used for any country to predict the vaccination rate when enough vaccine doses have been administered. The simulation can be used to determine which part of the society ignores vaccines more often than others. This model can also assist decision-makers in taking into account media advertisements that focus on the positive aspects of vaccines. The first part of the findings, including vaccine hesitancy and fully vaccinated prediction, validates the usefulness of the simulation as a platform to explore the effects of information diffusion, and these values correlate reasonably well with \cite{tang2021quantifying}. Moreover, the Monte Carlo test examined the model's robustness, and with a very minor non-alignment, it was found to be robust. The most remarkable result to emerge from the data is that information diffusion affects the vaccination rate, and modelers should consider this behavior in their model or simulation for vaccination. We believe that this is the first time that information diffusion is considered an effective factor in a vaccination simulation context. This study has gone some way towards enhancing our understanding of modeling vaccine acceptance. The present findings might help to solve the vaccine hesitancy problem in different countries. Our method can be applied to any country as long as the parameters are initialized correctly. Our data suggest that favorable information propagation in social networks could increase vaccine acceptance and speed up the vaccination rate. As our findings come from odd ratios reported by \cite{lunsky2021beliefs}, these findings should be treated with caution. We are aware that our research may have two limitations. The first is our regression model dependent on the reported odds ratios, and the second is that some important factors such as COVID-19 rate, vaccine effectiveness, and side effects are not well studied. These limitations highlight the difficulty of collecting data and the complexity of the model. 

\section*{Acknowledgement}
The research of AA, JK and JW has been funded by the Canadian Institute of Health Research (CIHR) 2019 Novel Coronavirus (COVID-19) rapid research program, and the Natural Science and Engineering Research Council of Canada's Emerging Infectious Disease Modelling Initiative (MfPH). JW is a member of the Ontario COVID-19 Modelling Consensus Table, sponsored by the Ontario Ministry of Health, Ontario Health, and Public Health Ontario. AA and JW are both members of the External Modelling Expert Panel of the Public Health Agency of Canada, and JW is a member of the Canada's Expert Pande on COVID Submittee on Modeling of the Chief Science Advisor of Canada.

\renewcommand\refname{References}
% This is the default "example" bibtex file which ships with the distribution. Be sure to
% comment out the following line if you want to disable citing all bibtex entries by
% default.
\bibliography{xampl.bib}
\newpage

\end{document}